\documentclass{elsarticle}

\usepackage{hyperref}

\usepackage{amssymb}
\usepackage{multirow}
\usepackage{graphicx}

\journal{Journal of Computational Science}

\begin{document}

\begin{frontmatter}

\title{Influence of Agents Heterogeneity in Cellular Model of Evacuation}
%
%

\author[vut]{Pavel Hrab\'{a}k\corref{cor1}}
\ead{pavel.hrabak@cvut.cz.cz}
\author[fjfi]{Marek Buk\'{a}\v{c}ek}
\ead{marek.bukacek@fjfi.cvut.cz}
\cortext[cor1]{Corresponding author. Tel.: +420 224 358 567; fax: +420 234 358 643}
\address[vut]{Brno University of Technology, Faculty of Civil Engineering, Veve\v{r}\'{i} 331/95, 602 00 Brno, Czech Republic}
\address[fjfi]{Czech Technical University in Prague, Faculty of Nuclear Sciences and Physical Engineering, Trojanova 13, 120 00 Prague, Czech Republic}

%
%
\begin{abstract}
The influence of agents heterogeneity on the microscopic characteristics of pedestrian flow is studied via an evacuation simulation tool based on the Floor-Field model. The heterogeneity is introduced in agents velocity, aggressiveness, and sensitivity to occupation. The simulation results are compared to data gathered during an original experiment.  The comparison shows that the heterogeneity in aggressiveness and sensitivity occupation enables to reproduce some microscopic aspects. The heterogeneity in velocity seems to be redundant.
\end{abstract}


\end{frontmatter}


\section{Introduction}

The presented study of the heterogeneity in CA models directly extends the contribution on aggressiveness presented in~\cite{HraBuk2016LNCS}. Detailed study of the heterogeneity in more aspects of agents in Floor-Field model is based on empirical observations related to variety of experiments conducted by our research group~\cite{BukHraKrb2015TGF,BukHraKrb2014Procedia,BukHraKrbTGF15}. We aim to investigate, whether some microscopic aspects of the pedestrian flow can be mimicked  introducing heterogeneity to some parameters of Floor-Field model. 

The experiments mentioned above were designed to study the boundary induced phase transition, which has been analysed theoretically in~\cite{EzaYanNis2013JCA} for Floor-Field model. The object of such study is a rather small room with one exit and one multiple entrance, which may be considered as one segment of a large network. During the experiment, an important aspect of pedestrian behaviour has been observed: participants have different ability to push through the crowd, what leads to significant variance in the time spent by the pedestrian in the room.

The presented cellular model is based on the Floor-Field Model~\cite{BurKlaSchZitPhysicaA2001,KirSch2002PhysicaA,Kretz2007PhD} with \emph{adaptive time-span}~\cite{BukHraKrb2014LNCS} and \emph{principle of bonds}~\cite{HraBukKrb2013JCA}. The adaptive time span enables to model heterogeneous stepping velocity of pedestrians; the principle of bonds helps to mimic collective behaviour of pedestrians in lines. For comprehensive summary of Floor-Field model modifications  capturing different aspects of pedestrian flow and evacuation dynamics we refer the reader to~\cite{SchChoNis2010}.

In this article we focus on the heterogeneity in three aspects of the pedestrian flow: the desired velocity, aggressiveness (or ability to win conflicts), and sensitivity to occupancy (related to line formation).

\section{Experiment}

\begin{figure}[h!]
	\begin{center}
	\hfill\includegraphics[height=.27\textwidth]{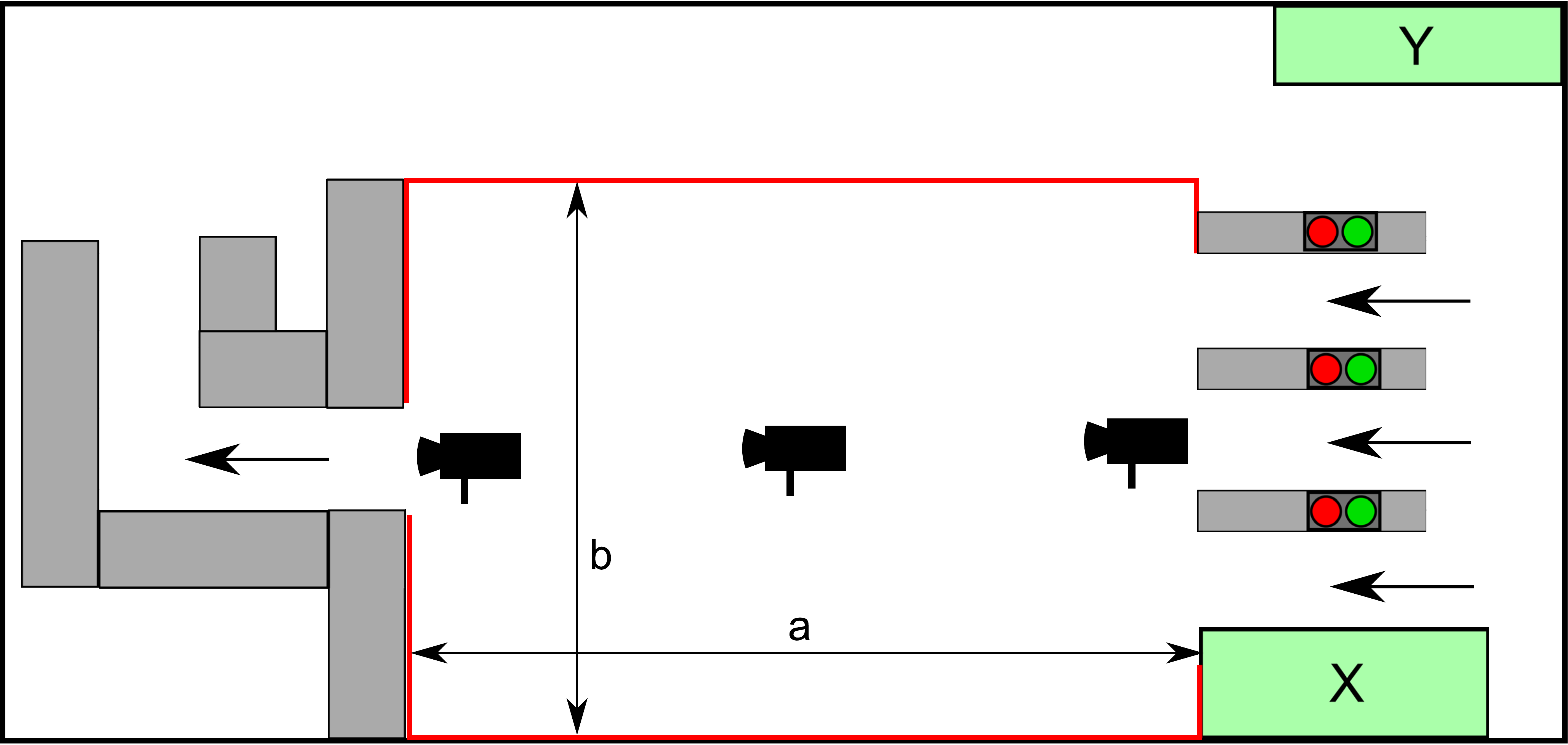}\hfill
	\includegraphics[height=.27\textwidth]{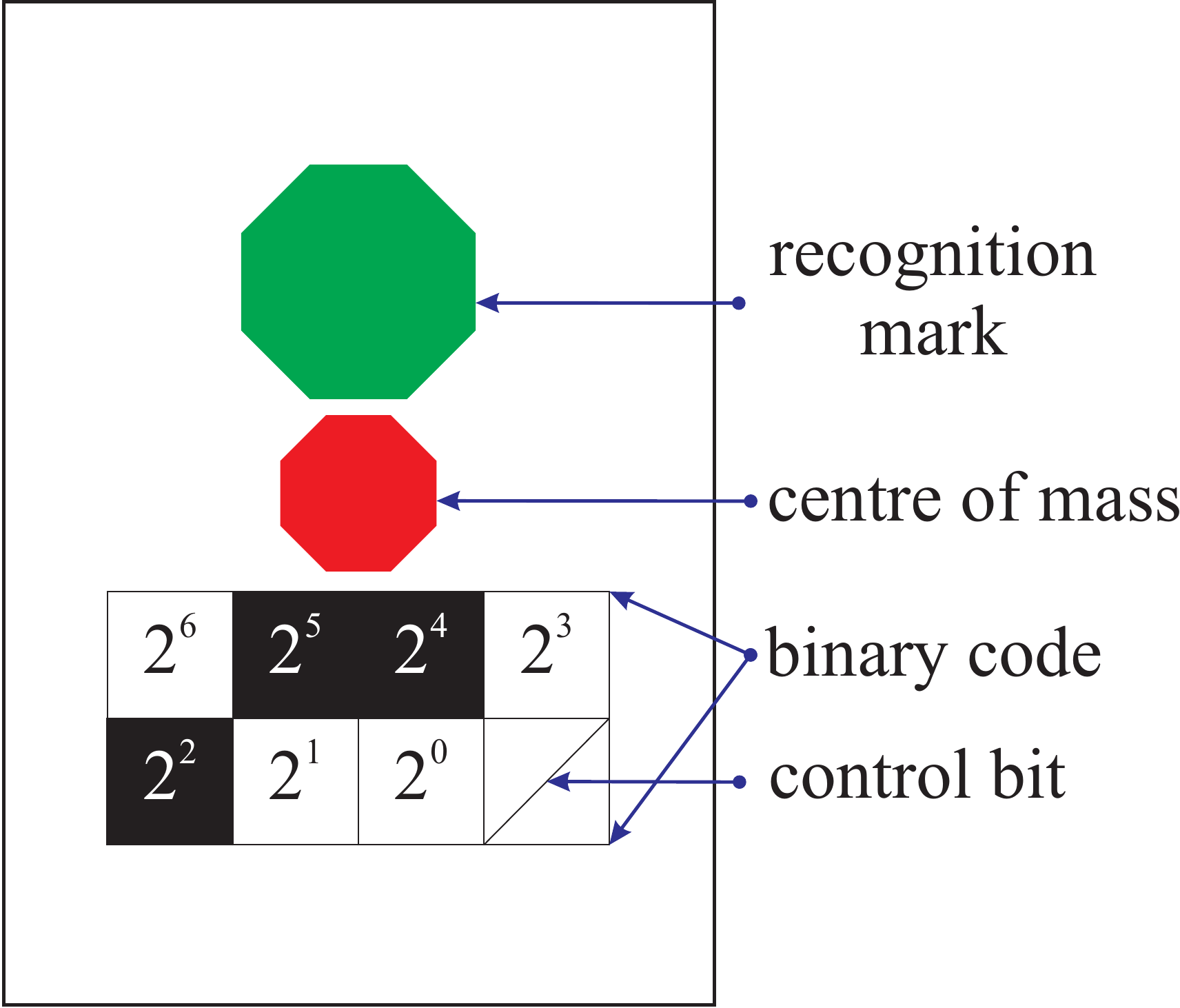}\hfill\phantom{x}
	\end{center}
\caption{Taken from~\cite{BukHraKrb2014Procedia}. Left: Setting of the experiment, a = 7.2~m, b = 4.4~m. Right: sketch of pedestrian's hat used for automatic image recognition.}
\label{fig:setting}
\end{figure}

Presented simulation study leans over the experiment ``passing-through'', which set-up is schematically depicted in in~Figure~\ref{fig:setting}. Detailed analyses of the experiment with respect to microscopic aspects of pedestrian flow can be found in~\cite{BukHraKrb2014Procedia,BukHraKrbTGF15}. Selected results of the above mentioned studies are summarized in this section. Videos capturing some aspects of the experiment are available at \url{http://gams.fjfi.cvut.cz/peds}.

In the experiment participants were instructed to walk through the rectangular room with one 60~cm wide exit and multiple entrance. The inflow of pedestrians $\alpha$~[ped/s] has been artificially controlled in order to investigate the boundary induced phase transition, which is not the subject of this article. Nevertheless, thanks to that the pedestrian behaviour could be observed under variety of conditions. The size of the crowd in front of the exit played the main role. During each run of the experiment, pedestrians were passing repeatedly through the room in order to keep stable inflow for sufficiently long time. Due to that there are 20 to 40 records for each participant. Moreover, the unique codes assigned to all participants enabled the study of individual properties of the pedestrians under variety conditions.

Usually the maximal outflow/capacity is related to bottleneck width $b$. In~\cite{SeyPasSteBolRupKli2009TS} the dependence is suggested $J_\mathrm{max}=1.9 \cdot b$. Here we note that the measured capacity of the bottleneck in the experiment was approximately 1.4~ped/s, which is higher than suggested. It was caused by high motivation of pedestrians to exit the room nad by the fact that the narrowing was fallowed by a slightly wider corridor. Despite that the motion of pedestrians within the following corridor was strictly one-lane without overtaking.

\subsection{Travel-Time}

The key investigated quantity is the travel time $TT$ denoting the time interval between the entrance at $T_\mathrm{in}$ and the egress at $T_\mathrm{out}$ of each pedestrian, i.e., $TT=T_\mathrm{out}-T_\mathrm{in}$. To capture the pedestrians behaviour under variety of conditions, the travel time is investigated with respect to the average number of pedestrians in the room $N_\mathrm{mean}$ defined as
\begin{equation}
\label{eq:Nmean}
	N_\mathrm{mean}=\frac{1}{T_\mathrm{out}-T_\mathrm{in}}\int_{T_\mathrm{in}}^{T_\mathrm{out}}N(t)\mathrm{d}t\,,
\end{equation}
where $N(t)$ stands for the number of pedestrians in the room at time $t$. As expected the average $TT$ increases with respect to the number of pedestrians in the room $N$, referred further as the occupancy. Surprisingly, the variance of the $TT$ increases dramatically with occupancy as well.

\begin{figure}[h!]
\centering
	\includegraphics[width=.9\textwidth]{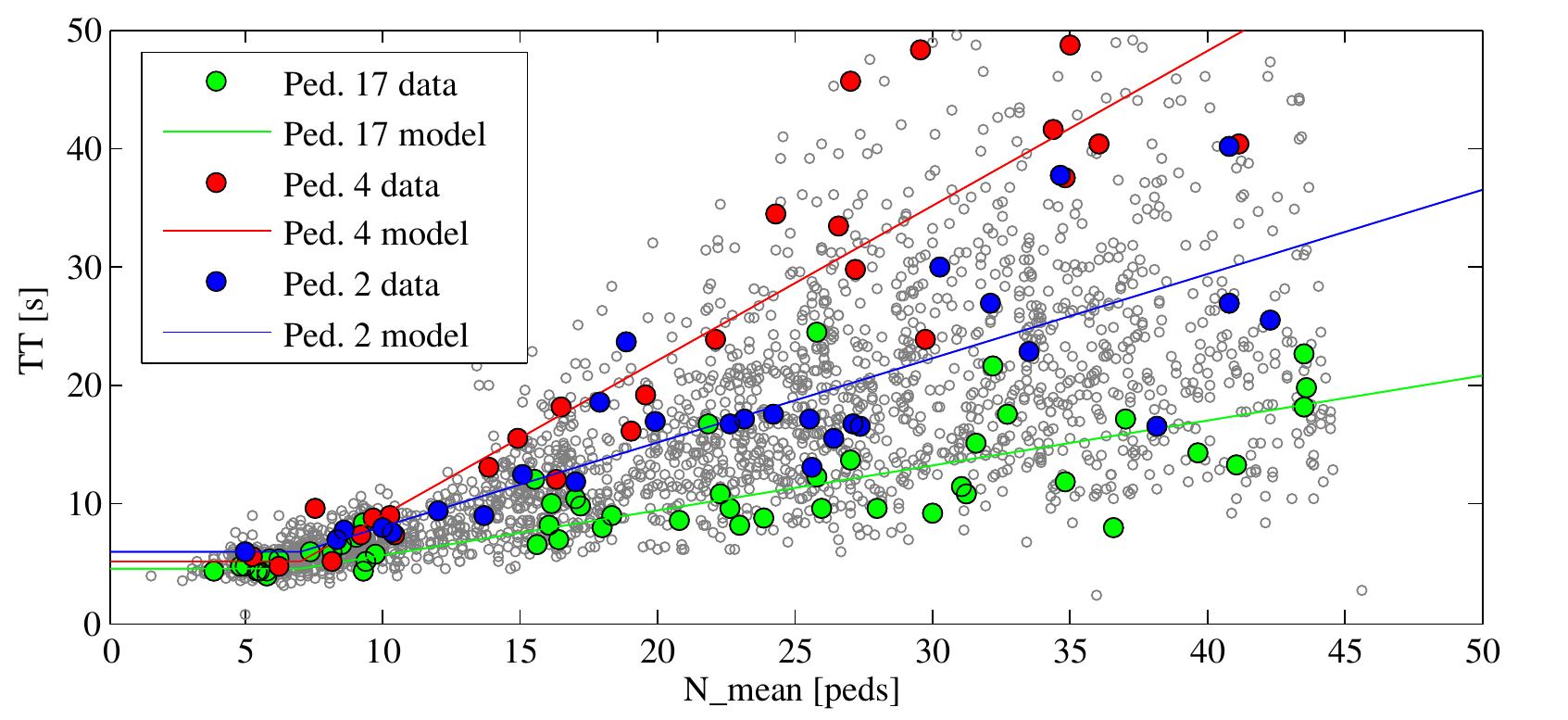}
\caption{Taken from~\cite{BukHraKrbTGF15}. Scatter plot of the travel time $TT$ with respect to the occupancy $N_\mathrm{mean}$ extracted from the experiment. Three participants are highlighted. Their travel time is approximated by the piecewise linear model~(\ref{eq:plm}). We can see that Ped. 2 has lower desired velocity in free regime but higher ability to push through the crowd in comparison to Ped. 4.}
\label{fig:TT-Nmean}
\end{figure}

Figure~\ref{fig:TT-Nmean} shows the scatter plot of all pairs $(N_\mathrm{mean},TT)$ gathered over all runs of experiment and all participants. Records corresponding to three chosen pedestrians are highlighted. We can observe that the reaction of participants to the occupancy $N$ significantly differs. The mean travel time in the free-flow regime (0 - 7 pedestrians) reflects the pedestrian desired velocity; the slope of the travel-time dependence on the occupation $N$ in the congested regime (10 - 45 pedestrians) reflects the pedestrian ability to push through or walk around the crowd. This observation corresponds to the piece-wise linear model for each pedestrian
\begin{equation}
\label{eq:plm}
	TT=\frac{S}{v_0(i)} + \mathbf{1}_{\{N>7\}}(N-7) \cdot \mathrm{slope}(i) + \mathrm{noise}\,
\end{equation}
where $S=7.2$~m, $v_0(i)$ is the free-flow velocity of the pedestrian $i$, $\mathrm{slope}(i)$ is the unique coefficient of the linear model for pedestrian $i$. The breakpoint $N=7$ depends from the room geometry. The weighted mean of the $R^2$ value of the model~(\ref{eq:plm}) is 0.688.

\subsection{Heterogeneity and Travel-Time}

\begin{figure}[h!]
\centering
	\includegraphics[width=.45\textwidth]{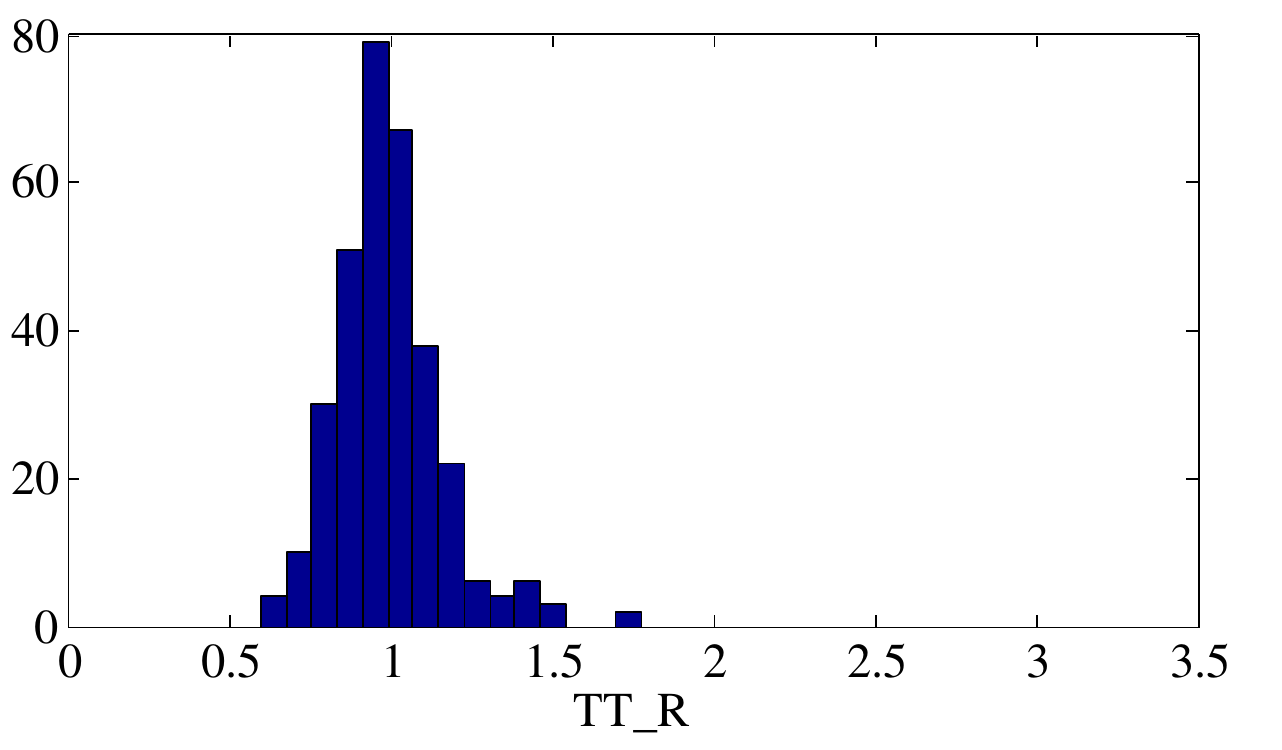}\hspace{.5cm}
	\includegraphics[width=.45\textwidth]{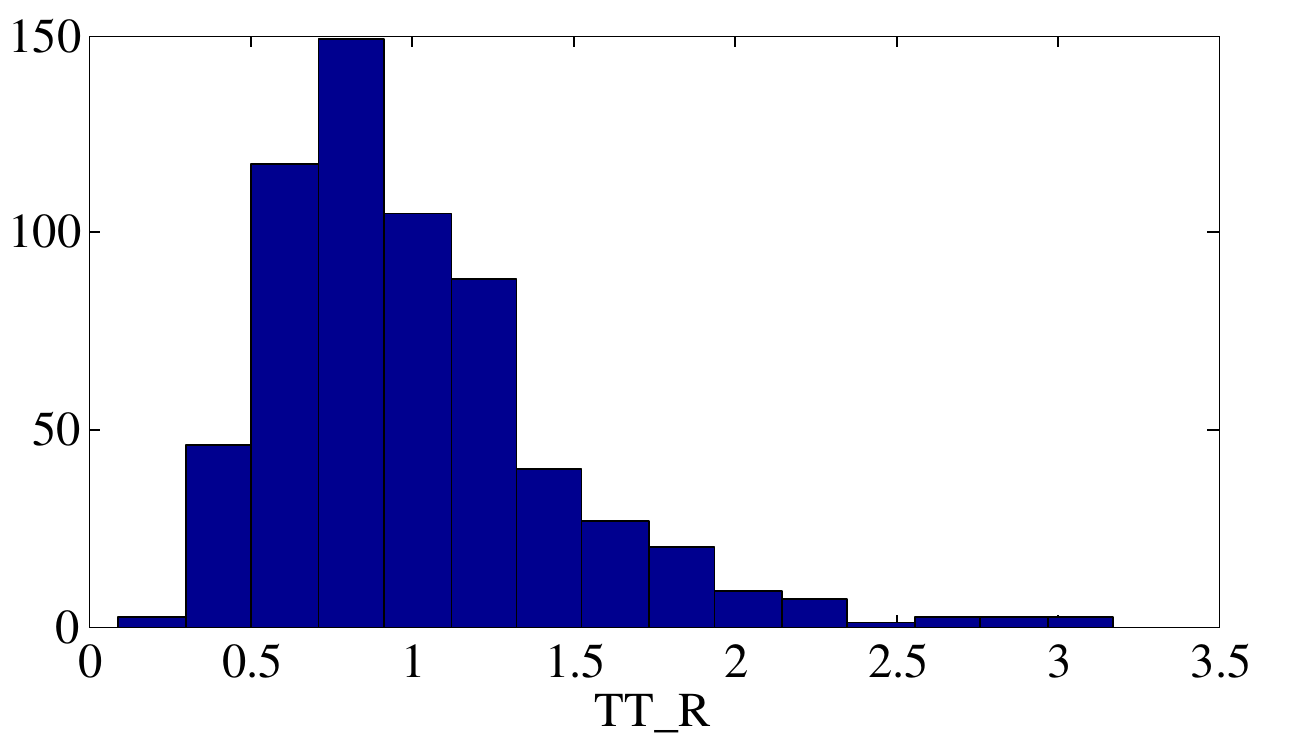}
\caption{Taken from~\cite{BukHraKrbTGF15}. Realative travel-time in free-flow 0-7~ped (left) and congested regime 30-45~ped (right).}
\label{fig:TTR}
\end{figure}

The high variance of the $TT$ caused by the heterogeneous reaction to crowd is reflected in the distribution of the relative travel-time $TT_R$, i.e., normalized $TT$ with respect to the average of records corresponding to similar occupancy $N_\mathrm{mean}$. Figure~\ref{fig:TTR} present the histograms of $TT_R$ for free flow regime (0-7 pedestrians) and for congested regime (30-50 pedestrians).

There are two aspects explaining the high variance of the $TT$ accompanied with extremely low and high values in congested regime: the aggressiveness and better path choice. We have observed that some of the pedestrians were pushing effectively through the crowd in order to get to the exit. This caused that some less aggressive pedestrians stayed trapped within the crowd or on its edge for significantly longer time.

\begin{figure}[h!]
\centering
	\includegraphics[width=4.5cm]{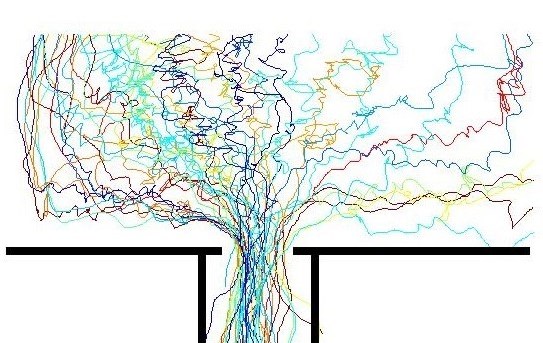}\hspace{1cm}
	\includegraphics[width=4.5cm]{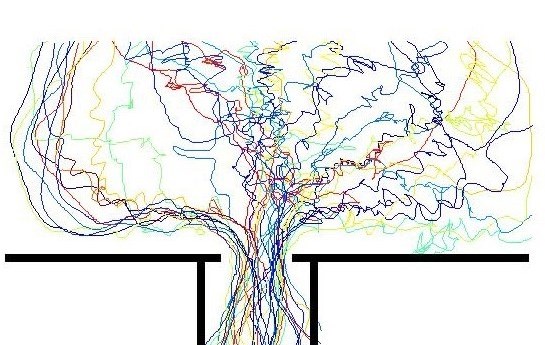}
\caption{Taken from~\cite{BukHraKrbTGF15}. Trajectories of pedestrians in front of the exit grouped according to the related travel-time, high $TT$ left and low $TT$ right.}
\label{fig:path}
\end{figure}

Furthermore, observing the path choice of ``fast'' and ``slow'' pedestrians we can make a conclusion that the lower travel-time was reached by walking around the crowd as well, see Figure~\ref{fig:path}.

\section{Model Definition}

The model used for this study comes out from the Floor-Field cellular model with several modifications introduced in~\cite{HraBukKrb2013JCA,BukHraKrb2014LNCS} and studied in more detail in \cite{BukHra2014LNCS} extended by the aggressiveness element~\cite{HraBuk2016LNCS}. The playground of the model is represented by the rectangular two-dimensional lattice $\mathbb{L}\subset\mathbb{Z}^2$ consisting of cells $x=(x_1,x_2)$. Every cell may be either occupied by one agent or empty. Agents are moving along the lattice by hopping from their current cell $x\in\mathbb{L}$ to a neighbouring cell $y\in\mathcal{N}(x)\subset\mathbb{L}$, where the neighbourhood $\mathcal{N}(x)$ is Moore neighbourhood, i.e.,
$
	\mathcal{N}(x)=\left\{y \in \mathbb{L};~\max_{j=1,2}|x_{j}-y_{j}|\leq1\right\}\,.
$

\subsection{Choice of the New Target Cell}

Agents choose their target cells $y$ from $\mathcal{N}(x)$ stochastically according to probabilistic distribution $P\left(y\mid x; \mathrm{~state~of~} \mathcal{N}(x)\right)$, which reflects the ``attractiveness'' of the cell $y$ to the agent. Part of the ``attractiveness'' is expressed by means of the static field $S$, where $S(y)=\sqrt{|y_{1}|^2+|y_{2}|^2}$ is the distances of cell $y\in\mathbb{L}$ to the exit cell $E=(0,0)$ acting as the common target for all agents. As usual,
$P(y \mid x)\propto\exp\{-k_SS(y)\}$, for $y\in\mathcal{N}(x)$, where $k_S\in[0,+\infty)$ denotes the parameter of sensitivity to the field $S$.

The probabilistic choice of the target cell is further influenced by the occupancy of neighbouring cells and by the diagonality of the motion. An occupied cell is considered to be less attractive, nevertheless, it is meaningful to allow the choice of an occupied cell while the principle of bonds is present (explanation of the principle of bonds follows below). Furthermore, the movement in diagonal direction is penalized in order to suppress the zig-zag motion in free flow regime and support the symmetry of the motion with respect to the lattice orientation.

Technically this is implemented as follows. Let $O_{x}(y)$ be the identifier of agents occupying the cell $y$ from the point of view of the agent sitting in cell $x$, i.e. $O_{x}(x)=0$ and for $y\neq x$ $O_{x}(y)=1$ if $y$ is occupied and $O_{x}(y)=0$ if $y$ is empty. Then $P(y\mid x)\propto(1-k_OO_x(y))$, where $k_O\in[0,1]$ is again the parameter of sensitivity to the occupancy ($k_O=1$ means that occupied cell will never be chosen). Similarly can be treated the diagonal motion defining the diagonal movement identifier as $D_{x}(y)=1$ if $(x_{1}-y_{1})\cdot(x_{2}-y_{2})\neq0$ and $D_{x}(y)=0$ otherwise. Sensitivity parameter to the diagonal movement is denoted by $k_D\in[0,1]$ ($k_D=1$ implies that diagonal direction is never chosen).

The probabilistic choice of the new target cell can be than written in the final form
\begin{equation}
\label{eq:Pxy}
	P(y\mid x)=\frac{\exp\big\{-k_{S}S(y)\big\}\big(1-k_{O}O_{x}(y)\big)\big(1-k_{D}D_{x}(y)\big)}{\sum_{z\in\mathcal{N}(x)}\exp\big\{-k_{S}S(z)\big\}\big(1-k_{O}O_{x}(z)\big)\big(1-k_{D}D_{x}(z)\big)}\,.
\end{equation}
It is worth noting that the site $x$ belongs to the neighbourhood $\mathcal{N}(x)$, therefore the equation~(\ref{eq:Pxy}) applies to $P(x\mid x)$ as well.

\subsection{Updating Scheme}
The used updating scheme combines the advantages of fully-parallel update approach, which leads to necessary conflicts, and the asynchronous clocked scheme~\cite{CorGreNew2005PhysicaD} enabling the agents to move at different rates.  

Each agent carries as his property the \emph{own period} denoted as $\tau$, which represents the minimal time between two steps, leading to desired times of update to be  $t=k\tau$, $k\in\mathbb{Z}$. The time-line is divided into isochronous intervals of the length $h>0$. During each algorithm step $k\in\mathbb{Z}$  such agents are updated, whose desired time of the next actualization lies in the interval $\big[kh,(k+1)h\big)$. A wise choice of the interval length $h$ in dependence on the distribution of $\tau$ enables to model heterogeneous velocity of agents while keeping sufficient number of conflicts. Here we note that we use the concept of adaptive time-span, i.e., the time of the desired actualization is recalculated after each update of the agent, since it can be influenced by the essence of the motion, e.g., diagonal motion leads to a time-penalization, since it is $\sqrt{2}$ times longer. For more detail see e.g.~\cite{BukHraKrb2014LNCS}. This is an advantage over the probabilistic approach introduced in~\cite{BanCroViz2015TGF}.

\subsection{Principle of Bonds and Sensitivity to Occupation}

The principle of bonds is closely related to the possibility of choosing an occupied cell.  An agent who chooses an occupied cell builds a bond to the agent sitting in there. This bond lasts until the motion of the blocking agent or until the next activation of the bonded agent. The idea is that the bonded agents attempt to enter their chosen cell immediately after it becomes empty. 

\begin{figure}
\centering
	\includegraphics[scale=1]{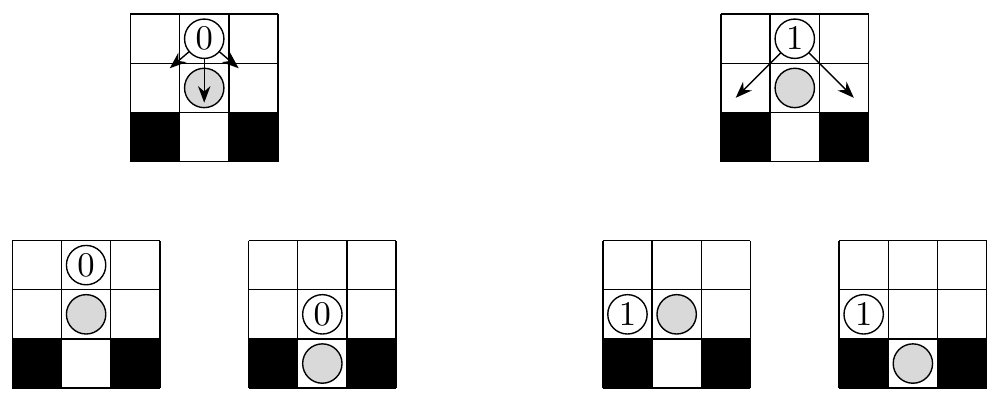}
\caption{Choice of the target cell with respect to sensitivity to occupancy. Left: $k_O=0$, bond is created. Right: $k_O=1$, motion to second most attractive cell.}
\label{fig:occupancy}
\end{figure}

The parameter $k_O$ (sensitivity to occupation) influences the willingness of agents to bond. The effect of different occupancy values is illustrated in Figure~\ref{fig:occupancy}. If $k_O=0$, agents choose the next target cell regardless whether it is occupied or empty, i.e, in case the most attractive cell (closer to the exit) is occupied, the agent creates a bond to the occupying agent assuming his motion within the step. This strategy supports the motion in lines and can lead to fluent flow through the bottleneck. On the other hand, the ignorance of occupancy leads to the increase of conflicts, which may block the motion in front of the exit. Here it is important to mention that the parameter $k_O<1$ influences the cell attractiveness before the normalization. Therefore, in case of extremely attractive cell (e.g. the exit), the probability distribution of choosing the next target cells does not change dramatically.

\subsection{Solution of Conflicts and Aggressiveness}

The in essence parallel updating scheme is inevitable accompanied by conflicts, when more agents decide to enter the same cell. In such case, one of the agents is usually chosen at random to win the conflict. There are more approaches, how the randomness is executed, e.g. uniformly, or proportionally to the hopping probabilities~\cite{BurKlaSchZitPhysicaA2001}. Important role in models of pedestrian evacuation play the unresolved conflicts, i.e., the aim to attempt the same cell leads to the blocking of the motion. This is captured by the friction parameter $\mu$ denoting the probability that none of the agent wins the conflict. An improvement is given by the friction function~\cite{YanKimTomNisSumOhtNish2009PRE}, which raises the friction according to the number of conflicting agents.

For purposes of this article, we introduce the choice of the winning agent based on his ability to win conflicts represented by an additional parameter $\gamma\in[0,1]$, which is referred to as the aggressiveness. Similar heterogeneity in agents behaviour has been used in~\cite{JiZhouRan2013PhysicaA}, where the ``aggressiveness'' has been represented by the willingness to overtake.

\begin{figure}
\centering
	\includegraphics[scale=1]{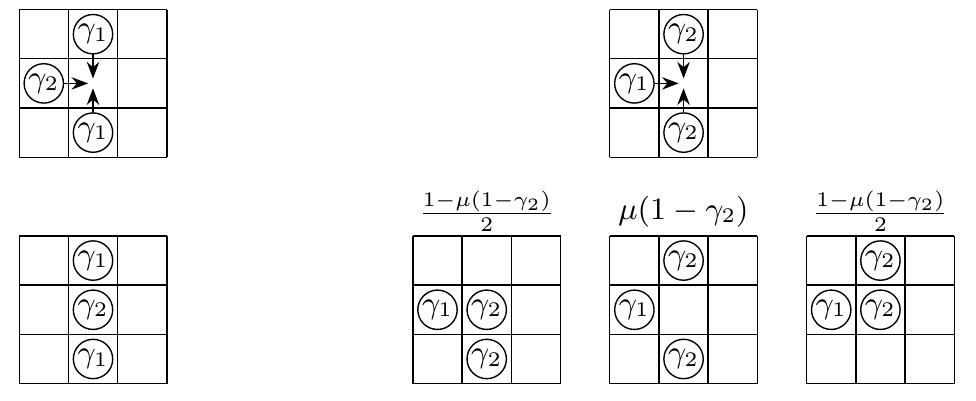}
\caption{Conflict solution for $\gamma_1<\gamma_2$. Left: More aggressive wins the conflict over two less aggressive. Right: The conflict of two more aggressive can resolve by the blocking of the movement.}
\label{fig:conflict}
\end{figure}

The conflict is always won by agents with highest $\gamma$ among conflicting agents. If there are two or more agents with the highest $\gamma$, the friction parameter $\mu$ plays a role. In this article we assume that the higher is the aggressiveness $\gamma$, the less should be the probability that none of the agents wins the conflict. Therefore, the conflict is not solved with probability $\mu(1-\gamma)$ (none of the agents move). With complement probability $1-\mu(1-\gamma)$ the conflict resolves to the motion of one of the agents. This agent is chosen randomly with equal probability from all agents involved in the conflict having the highest $\gamma$. The conflict solution among heterogeneous group of agents is depicted in Figure~\ref{fig:conflict}.

\section{Simulation Setting}

Main goal of this article is to study the influence of heterogeneity in chosen parameters. In this article we focus on the heterogeneity in free-flow velocity (represented by own frequency), aggressiveness, and sensitivity to occupation. In order to have comparable results to the experiment, the parameters have been calibrated to give similar values of important macroscopic quantities as free flow velocity (1.57~m/s in experiment) and maximal outflow (1.4~ped/s in experiment). The used set of parameters is given in Table~\ref{tab:parval}.

\begin{table}[h!]
\caption{Parameter values and description}
\label{tab:parval}
\centering
	\begin{tabular}{l|lll}
		\hline
			Parameter & Value & Range & Description\\
		\hline
			$\Delta x$ [cm] & $40$ & & Lattice constant\\
			$k_S$ & 3.5 & $[0,\infty)$& Sensitivity to potential\\
			$k_D$ & 0.7 & $[0,1]$ & Penalization of diag. motion\\
			$\mu$ & 0.9 & $[0,1]$& Friction parameter\\
			$h$~[s] & 0.2 &$(0,\infty)$ & Length of one time step\\
			$\tau$~[s]& 0.2 & $(0,\infty)$ & Homogeneous own period\\
			& $\{0.15, 0.4\}$ & & Heterogeneous own period\\
			$\gamma$ & 0.14 & $[0,1]$ & Homogeneous aggressiveness\\
			& $\{0,1\}$ & & Heterogeneous aggressiveness\\
			$k_O$ &0.9& $[0,1]$ & Homog. sensitivity to occupation\\
			 &$\{0.1,0.95\}$&  & Heter. sensitivity to occupation\\			
		\hline
	\end{tabular}
\end{table}

The free flow simulation (without interactions) is directly influenced by $k_S$, $k_D$ and $\tau$. The diagonal penalization $k_D$ together with time penalization of diagonal motion have been tested in previous research. The values of $k_S$ and $\tau$ have been chosen to agree with the mean and variance of the free-flow velocity. Here we note that the pedestrians in the experiment walked relatively fast (1.57~m/s), which motivated us to set the algorithm step $h=0.2$~s of real time to balance significant decrease of velocity in congested regime.

The motion of agents in crowd is influenced by parameters $\mu$, $\gamma$, and $k_O$. These parameters have been calibrated by means of the maximal outflow from the exit in congested situation, i.e, the exit capacity (1.4 ped/s). The significant decrease of velocity in crowd is modelled by means of relatively high friction $\mu=.9$. Here we note that such high friction is necessary to compensate the conflict solution mechanism related to aggressiveness and the motion in lines related to the sensitivity to occupation.

\begin{figure}[]
\centering
	\includegraphics[width=.9\textwidth]{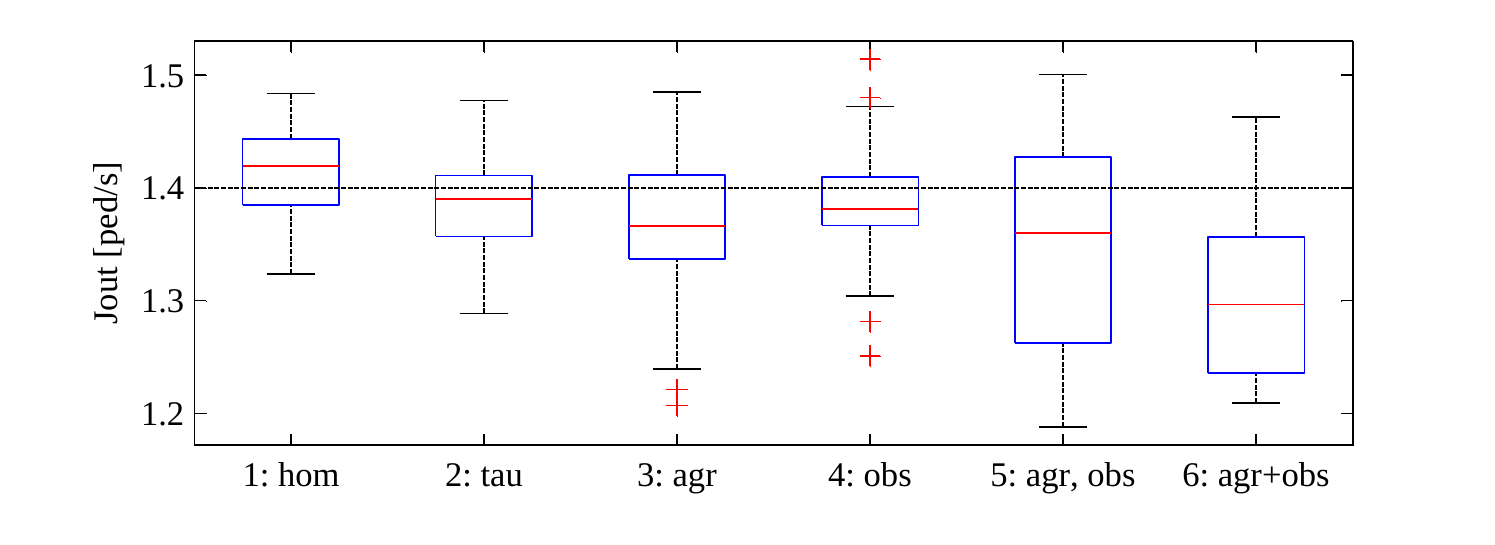}
\caption{The box-plots of outflow $J_\mathrm{out}$ from the congested room ($N=50$) measured for 20 runs of the simulation experiment for each parameter set.}
\label{fig:jout}
\end{figure}

Our goal is to illustrate the effects of heterogeneity in chosen parameters. Therefore, the values of outflow had not been calibrated directly to the value 1.4~m/s, but sufficiently close to it, see Figure~\ref{fig:jout}. Such approach enables to fit the homogeneous and heterogeneous values of each parameter independently and therefore can be used in all considered scenarios -- 1: hom (homogeneous in all parameters), 2: tau (heterogeneous in velocity), 3: agr (heterogeneous in aggressiveness), 4: obs (heterogeneous in sensitivity to occupation), 5: agr,obs (heterogeneous in aggressiveness and occupation independently), 6: agr+obs (heterogeneous in aggressiveness and occupation with dependence, i.e., more aggressive is more sensitive to occupation). 

The simulation has been performed for system with periodic boundaries, i.e., the egress of an agent causes the entrance of another one (contrary to~\cite{HraBuk2016LNCS}, where open boundaries were used giving the same results). The properties of agents $(\tau,\gamma,k_O)$ were drawn from uniform distribution on the groups of parameters. For each set of parameters the simulation was performed for the occupancy $N\in\{1, 3, 5, 7, 10, 12, 14, 17, 20, 30, 40, 45, 50, 75, 100\}$. The simulation was performed until 1000 agents walked through the exit and it was repeated 20 times. All quantities are averaged. 

\section{Simulation Results}

In Table~\ref{tab:avrgquan} the measured values of free flow velocity $v_0$ (velocity for agents under the occupancy $N\leq4$), maximal outflow $J_\mathrm{out}$, and average travel-time $\overline{TT}_N$ for all 6 settings are compared to the experimental values. From the values we can see that the average travel-time is slightly, but not significantly, overestimated by the model. The lower free-flow velocity in case 2: tau is caused by the heterogeneity in velocity. The used set of parameters did not allow to fit both, the free-flow velocity and the outflow in this case. The reason lies in the synchronous update with $h=0.2$~s. As will be discussed below, the heterogeneity in velocity defined by the own frequency is not desirable for the presented model with respect to the performed experiment.

\begin{table}[h!]
\caption{Average quantities measured for different parameter settings.}
\label{tab:avrgquan}
\centering
	\begin{tabular}{l|rrrrrrr}
		\hline
			 & Exp. & 1:hom & 2:tau & 3:agr & 4:obs & 5:{\small agr,obs} & 6:{\small agr+obs}\\
		\hline
			$v_0$~[m/s]& 1.57 & 1.11 & 1.57 & 1.57 & 1.57 & 1.57 & 1.57 \\
			$J_\mathrm{out}$~[ped/s]& 1.40 & 1.42 & 1.39 & 1.37 & 1.38 & 1.35 & 1.30\\
			$\overline{TT}_{45}$~[s]& 24.31 & 30.74 & 30.76 & 30.72 & 31.17 & 32.01 & 33.05\\
			$\overline{TT}_{100}$~[s]& -- & 67.74 & 66.83 & 67.84 & 67.52 & 67.63 & 70.59\\			
		\hline
	\end{tabular}
\end{table}

\begin{figure}
\centering
	\includegraphics[width=\textwidth]{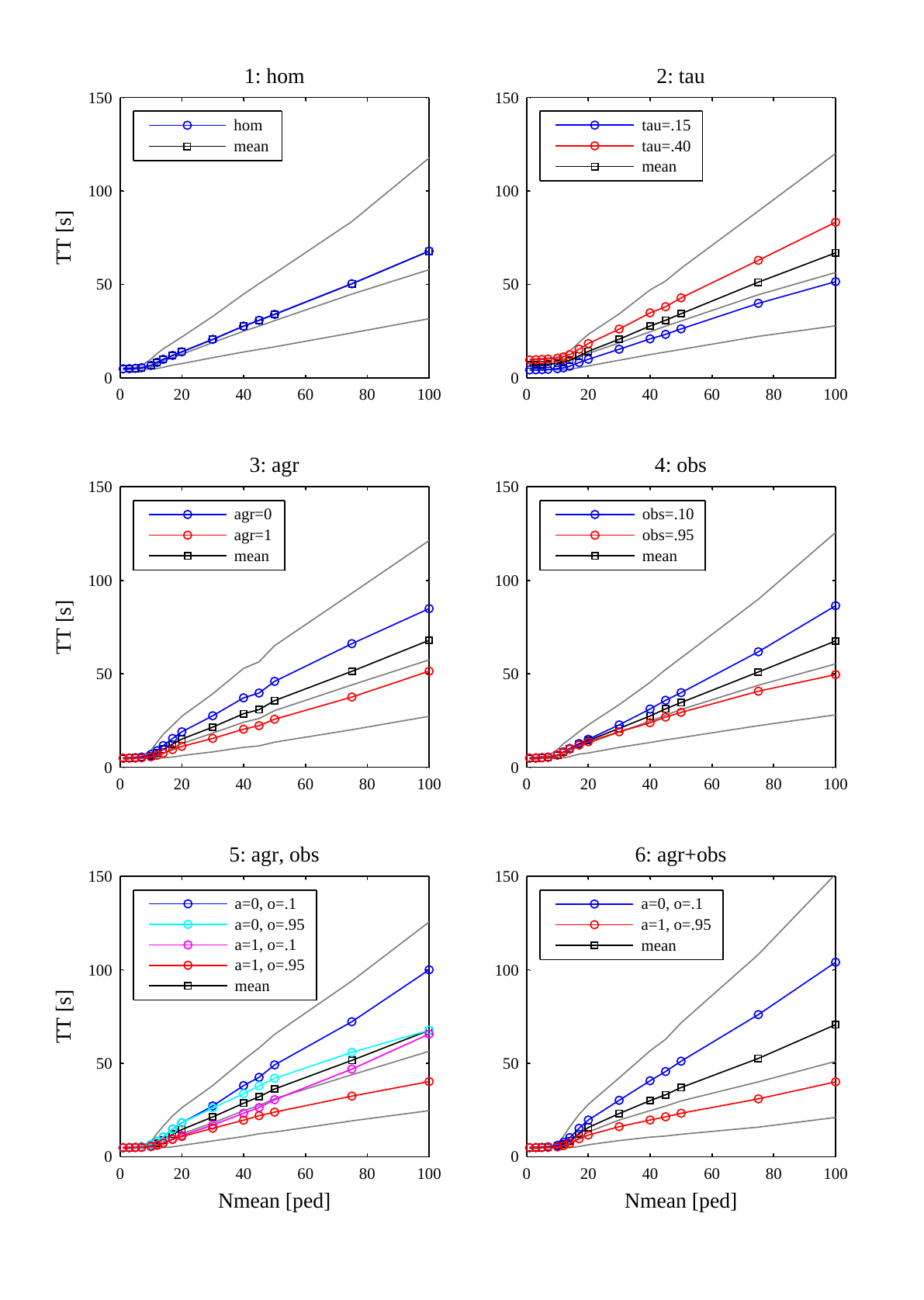}
\caption{Average travel against mean occupancy plotted by groups. Black line represents the mean over all entries, gray lines correspond to 0.1, 0.5, and 0.9 quantils. }
\label{fig:TTsim}
\end{figure}

The main stress is given to the travel-time study and dependence on the average occupancy $N_\mathrm{mean}$, which reflects the heterogeneity in the reaction to the crowd. The $TT-N_\mathrm{mean}$ plots for studied parameter sets are given in Figure~\ref{fig:TTsim}. In the graphs the mean travel time is plotted according by groups with the same parameters, the overall mean and quantiles are present for completeness. From the graphs is evident that the model is able to mimic the piecewise linear dependence~(\ref{eq:plm}) of $TT$ on $N_\mathrm{mean}$, which is present in the experimental study. The break-point of the model is in agreement with the experimental observation $N=7$.

We can see quite good agreement with the experiment regarding the trend of the dependence of $TT$ on the occupancy not only in average but also in the minimal and maximal measured values, see Figure~\ref{fig:TT-Nmean}. Due to to significantly lower volume of data from the experiment it is reasonable to compare the 0.1 and 0.9 quantiles. Here we note that the lower average $TT$ in experiment (24.31~s) than produced by the model (approx. 30~s) is caused  by the small volume of data related to $N\approx45$ in the experiment -- such crowded conditions were kept for relatively short time due to the small number of participants (75).

We can see that the differences in the slope of the linear dependence can be observed in all heterogeneous scenarios. Nevertheless, for further studies we have neglected the heterogeneity in the own updating period $\tau$, related to velocity. Even the homogeneous setting of free-flow velocity can reproduce variances in the free-flow travel-time in sufficient manner due to the stochastic nature of Floor-Field model. Therefore, another heterogeneity in the parameter $\tau$ is redundant. On the other hand, the concept can be used in case of evident heterogeneity where the desired velocity significantly differs~\cite{KleWas2014LNCS,SpaGeoSir2014LNCS}.

The heterogeneity in aggressiveness parameter $\gamma$ becomes evident in occupancy $N\in(10,50)$, the effect does not rise significantly for higher occupancy, see sub-figure 3. Complementary evinces itself the heterogeneity in sensitivity to occupancy $k_O$ (sub-figure) 4, which becomes most evident for occupancy $N>50$. Interesting results brings the combination of these two parameters. The scenario 5 shows that the heterogeneity in $\gamma$ and $k_O$ combines both effects from 3 and 4 in superposition manner. The main differences in the slope of $TT$ shows the scenario 6, where the aggressiveness and occupancy are dependent, i.e., the more aggressive agents are more sensitive to occupation.

Here we note that the sensitivity to occupation should be interpreted as the willingness to join an existing queue in order to move along this queue. It is therefore reasonable to suppose that the aggressive behaviour is related to the willingness to overtake, i.e., $k_O\to1$

\begin{figure}[h!t]
\centering
	\includegraphics[width=\textwidth]{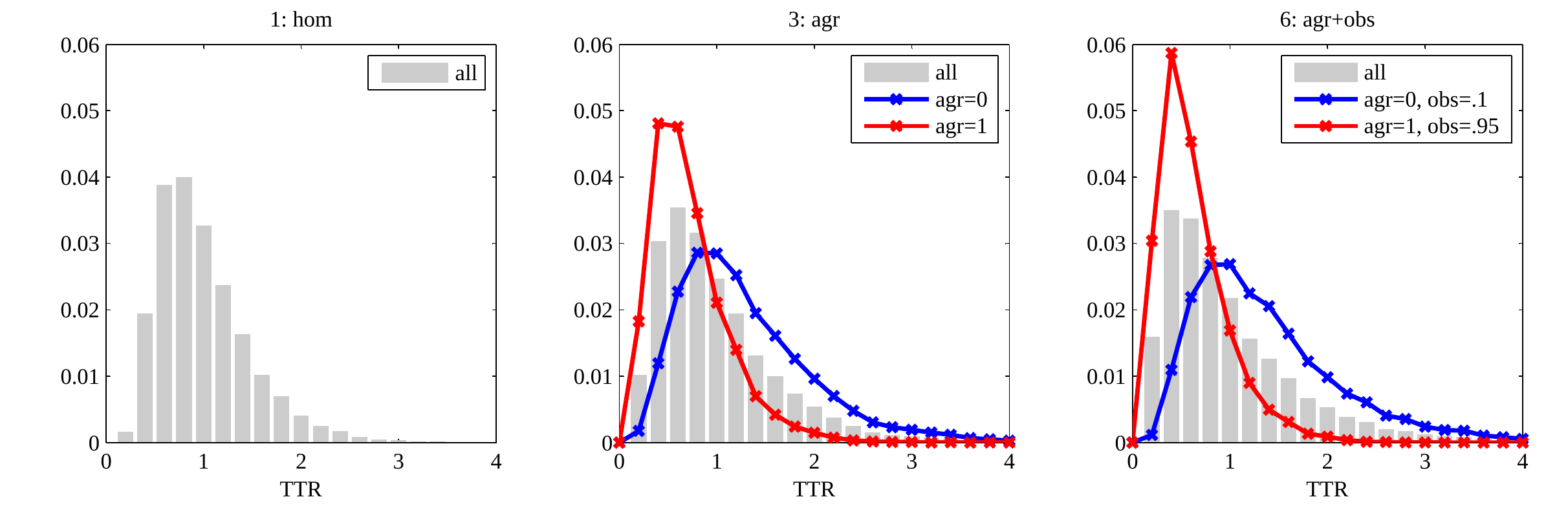}
\caption{normalized histograms of relative travel-time $TT_R$ for scenarios 1: hom, 3: agr, and 6: agr+obs.}
\label{fig:TTRsim}
\end{figure}

For further comparison let us focus on the distribution of the relative travel time $TT_R$. In Figure~\ref{fig:TTRsim} we can see the histograms of $TT_R$ for scenarios 1, 3, and 6, i.e., with increasing heterogeneity. We can make a conclusion that with increasing heterogeneity increases the relative frequency of low values of $TT_R$ (first bin). Similarly, the modus of the distribution is closer to lower values for higher heterogeneity (bins 4, 3, 2).

Looking at the histograms of $TT_R$ related to heterogeneous scenarios, we conclude that the final distribution can be considered as a mixture of two uni-modal distribution corresponding to two groups of agents, which seems to be the case of the experiment as well, see Figure~\ref{fig:TTR}.

\begin{figure}[h!t]
\centering
	\includegraphics[width=.25\textwidth]{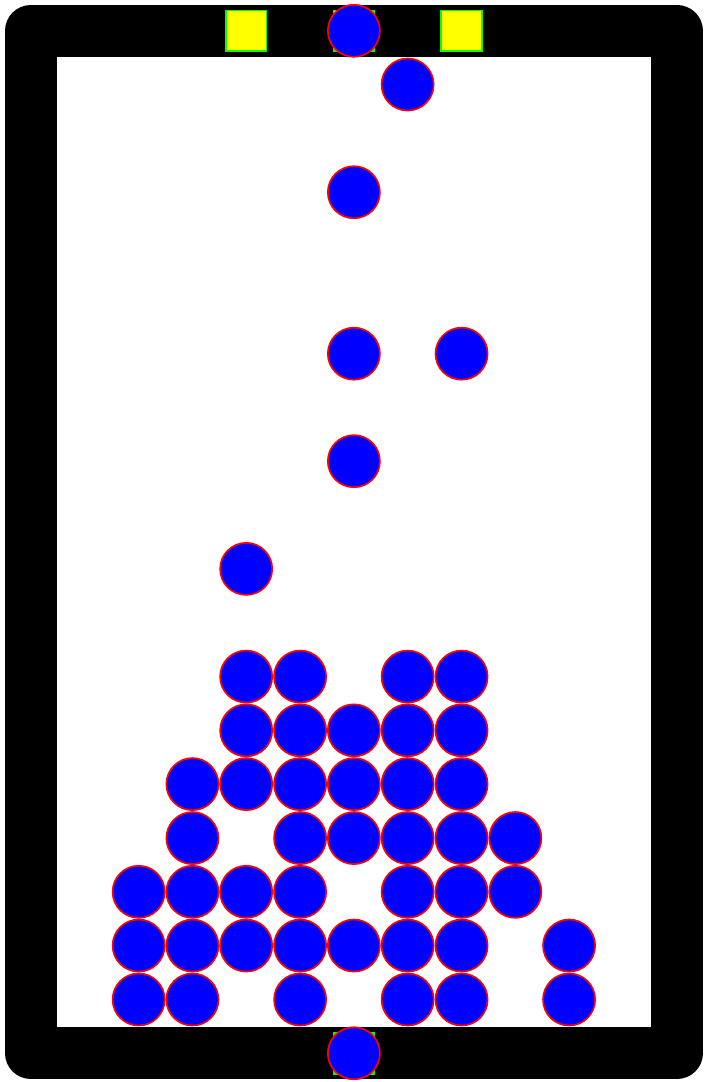}\hspace{1cm}
	\includegraphics[width=.25\textwidth]{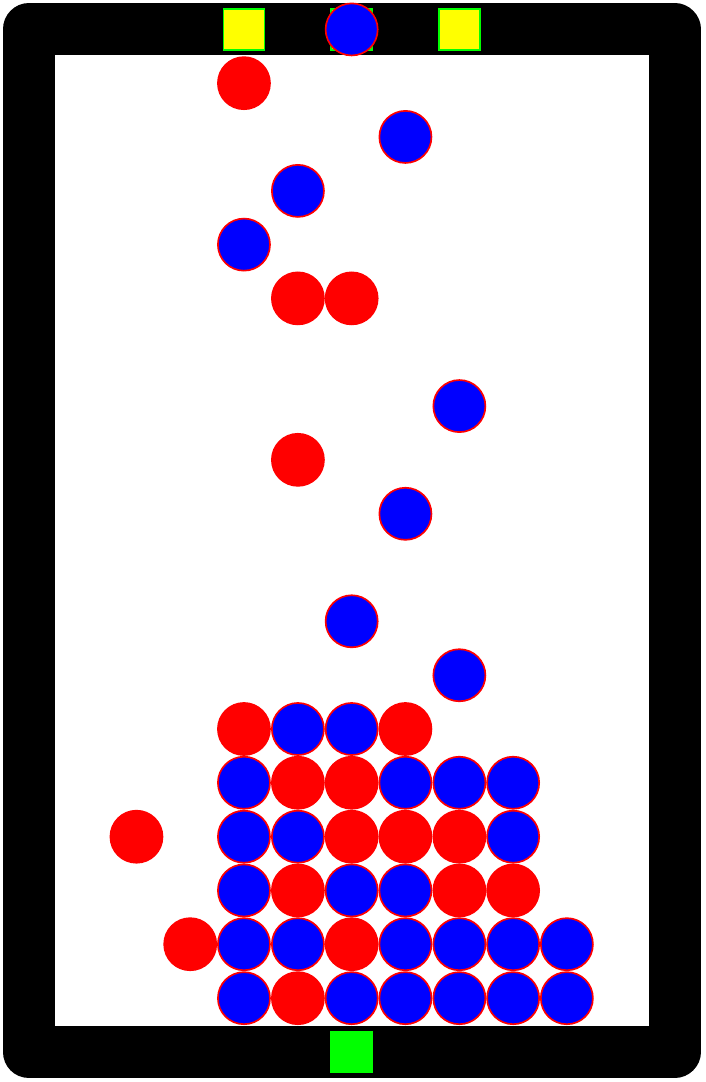}\hspace{1cm}
	\includegraphics[width=.25\textwidth]{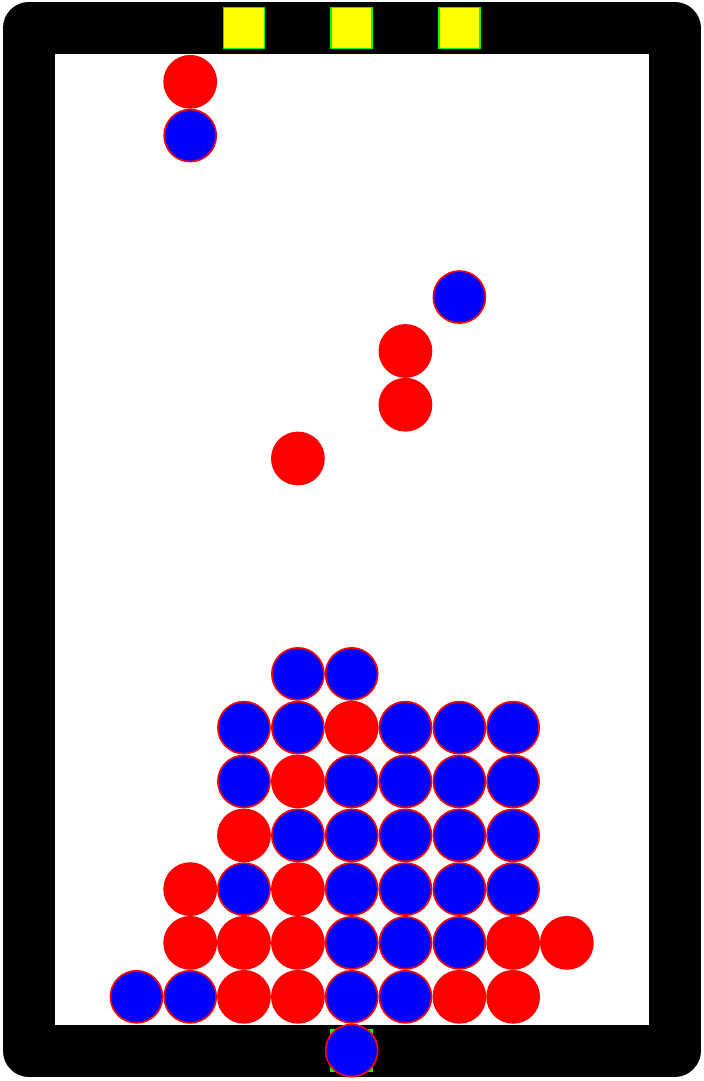}
\caption{Snapshots from the simulation of scenarios 1: hom, 3: agr, and 6: agr+obs approximately 60~s after initiation. By red dots are marked more aggressive pedestrians.}
\label{fig:pathchoice}
\end{figure}

Although some differences between scenarios 3 and 6 six are evident, they might be considered as marginal in comparison with the homogeneous case 1. Let us therefore focus on another aspect observed during the experiment -- the path choice. In Figure~\ref{fig:pathchoice} we can see the snapshots from the simulation showing a representing situation of the simulation. In scenario 3, the more aggressive agents are more successful in pushing through the crowd, but they do not evince any preference in path-choice. In scenario 6, the aggressive agents prefer walking around the crowd and hopping to the exit from the left or right. As a consequence, less aggressive agents standing in lines remain often trapped few cell away from the exit, as often observed during the experiment.

\section{Summary and Conclusions}

The article focuses on the heterogeneity in three parameters of the original cellular model based on Floor-Field model. These parameters are then considered as the properties carried by individual agents, specifically the heterogeneity in own period (velocity), aggressiveness, and sensitivity to occupancy is studied. It is important to mention that the introduced concept of aggressiveness and principle of bonds can be used in other rule-based cellular model of pedestrian dynamics.

The introduction of heterogeneity into the model has been inspired by the results of conducted experiments. It has been show that pedestrians react differently to the crowd: some are able to push through the crowd effectively, some remain trapped in front of the exit. Moreover, this is not a property of individual trajectories, but this ability is closely related to the pedestrians.

In this study we aim to show that the heterogeneity in the ability to win conflicts is necessary to reproduce the above mentioned effects in cellular models. Such property can be useful for proper modelling of an evacuation of a complex structure, where the heterogeneity in the ability to win conflicts is expected. In such cases a group of people can remain trapped within the facility for unjustifiable long time, although the average flows and evacuation times fulfil the expectations.

The introduced aspects of heterogeneity can be summarized as follows:

\begin{enumerate}
	\item \textbf{Velocity:} The heterogeneity in velocity causes undesired bi-modal histogram in the free-flow regime. The observed heterogeneity in pedestrian sample can be sufficiently modelled by the stochastic nature of the decisions, the variance in the travel time is then related to the number of deviations from the direct path. Nevertheless, the concept can be used in dramatically heterogeneous scenario where the average speed of one group of pedestrians is two times higher than the average velocity of other group.
	\item \textbf{Aggressiveness:} This conflict-solution method, where the conflict is won by the agent with higher value of aggressiveness, seems to be very effective in the reproduction of high variance of the $TT$ in the congested regime. In combination with the heterogeneity of the velocity we are able to simulate a situation when slower agent is more aggressive than a fast one. It is important to note that the term aggressiveness may be a little bit misleading, since it corresponds to the ability to win conflicts, which may be given by some rules of preference as well. 
	\item \textbf{Sensitivity to Occupancy:} This aspect influences mainly the space usage by the agent in given conditions. The lower the sensitivity is, the higher is the average $TT$ for the agent, since he waits in lines and can be overtaken and trapped in front of the exit. This parameter plays very important role in combination with the parameter of aggressiveness.
\end{enumerate}

\section*{Acknowledgements}
This work was supported by the Czech Science Foundation grant GA15-15049S and by the Czech Technical University grant SGS15/214/OHK4/3T/14.

\section*{References}

\bibliography{JCS_PH+MB_references}

\end{document}